\documentclass{appolb}
\usepackage{epsfig}
\usepackage{amsmath}
\newcommand\figref{Figure~\ref}

\def\sr17{$\sqrt{s}$~=~17~GeV~}

\newcommand{\be}{\begin{equation}}
\newcommand{\ee}{\end{equation}}

\begin{document}
\newcount\eLiNe\eLiNe=\inputlineno\advance\eLiNe by -1
\title{Transverse Momentum Distributions at the LHC and Tsallis Thermodynamics.}
%
%
\author{M. D. Azmi and J. Cleymans
\address{UCT-CERN Research Centre and Department of Physics\\
University of Cape Town,
Rondebosch, South Africa 
	  }}

\maketitle
\begin{abstract}
An overview is presented of  
transverse momentum distributions of particles at the LHC using the Tsallis distribution. 
The use of a thermodynamically consistent
form of this distribution leads to an excellent description of charged and identified particles.
The values of the Tsallis parameter $q$ are truly remarkably consistent.
\end{abstract}
\section{\label{sec:Introduction}Introduction}
It is by now standard to parameterize transverse momentum distributions
with functions having a power law behaviour at high momenta. This has been done 
by the  STAR~\cite{STAR} and PHENIX~\cite{PHENIX} collaborations at RHIC and by the  
ALICE~\cite{ALICE}, ATLAS~\cite{ATLAS} and CMS~\cite{CMS} collaborations at 
the LHC.
In this talk we would like to pursue the use of the Tsallis distribution to
describe transverse momentum distributions at the highest beam energies. \\
\indent In the framework of Tsallis 
statistics~\cite{tsallis,biro,miller,worku1,worku2}
the entropy $S$, the particle number, $N$,  the energy density $\epsilon$ and the pressure $P$ 
are given by corresponding
integrals over the  Tsallis distribution: 
\begin{equation}
f = \left[ 1 + (q-1) \frac{E-\mu}{T}\right]^{-\frac{1}{q-1}} .\label{tsallis} 
\end{equation}
\indent It can be shown (see e.g.~\cite{worku2}) that the relevant thermodynamic quantities are given by:
\begin{eqnarray}
	S&=&-gV\int\frac{d^3p}{(2\pi)^3}
\left[ f^{q}\ln_{q} f -f \right],
\label{entropy} \\
N &=& gV\int\frac{d^3p}{(2\pi)^3} f^q ,\label{number} \\
\epsilon &=& g\int\frac{d^3p}{(2\pi)^3}E f^q ,\label{epsilon}\\
P &=& g\int\frac{d^3p}{(2\pi)^3}\frac{p^2}{3E} f^q\label{pressure} .
\end{eqnarray}
where $T$ and $\mu$ are the temperature and the chemical potential,
$V$ is the volume and  $g$ is the degeneracy factor.  
We have used the short-hand notation
\begin{equation}
\ln_q (x)\equiv \frac{x^{1-q}-1}{1-q} , \label{suba} 
\end{equation}
often referred to as q-logarithm.
It is straightforward to show that the relation
\begin{equation}
	\epsilon + P = Ts + \mu n
\end{equation}
(where $n, s, \epsilon$ refer to the densities of the corresponding quantities)
is satisfied.
The first law of thermodynamics gives rise to the following  
differential relations:
\begin{eqnarray}
 d\epsilon = Tds + \mu dn,\label{a5}\\
dP = sdT + nd\mu.\label{a51}
\end{eqnarray}
\indent Since these are total differentials, thermodynamic consistency requires the following
Maxwell relations to be satisfied:
\begin{eqnarray}\label{a6}
 T &=& \left.\frac{\partial \epsilon}{\partial s}\right|_{n},\label{a61}\\
 \mu &=&\left.\frac{\partial \epsilon}{\partial n}\right|_{s},\\
 N &=& V\left.\frac{\partial P}{\partial \mu}\right|_{T},\label{a63}\\
 S &=& V\left.\frac{\partial P}{\partial T}\right|_{\mu}.\label{a64}
\end{eqnarray}
\indent This is indeed the case, e.g. for Eq.~\eqref{a63} this follows from
\begin{eqnarray}
\left.\frac{\partial P}{\partial \mu}\right|_{T}
&=&- g\int\frac{d^3p}{(2\pi)^3}\frac{p^2}{3}
\frac{d}{EdE}\left[ 1 + (q-1) \frac{E-\mu}{T}\right]^{-\frac{q}{q-1}}
\nonumber \\
&=&- g\int\frac{d^3p}{(2\pi)^3}\frac{p^2}{3}
\frac{d}{pdp}\left[ 1 + (q-1) \frac{E-\mu}{T}\right]^{-\frac{q}{q-1}}
\nonumber \\
&=& g\int\frac{d\cos\theta d\phi dp}{(2\pi)^3} 
\left[ 1 + (q-1) \frac{E-\mu}{T}\right]^{-\frac{q}{q-1}}
\frac{d}{dp}\frac{p^3}{3}
\nonumber \\
&=& n
\nonumber
\end{eqnarray}
after an integration by parts and using  $p~dp = E~dE$.\\
\indent Following from Eq.~\eqref{number}, the momentum distribution is given by:
\begin{equation}
\frac{d^{3}N}{d^3p} = 
\frac{gV}{(2\pi)^3}
\left[1+(q-1)\frac{E -\mu}{T}\right]^{-q/(q-1)},
\label{tsallismu}
\end{equation}
or, expressed in terms of transverse momentum, $p_T$,  
the transverse mass, $m_T \equiv \sqrt{p_T^2+ m ^2}$, and the rapidity  $y$  
\begin{equation}
\frac{d^{2}N}{dp_T~dy} = 
gV\frac{p_Tm_T\cosh y}{(2\pi)^2}
\left[1+(q-1)\frac{m_T\cosh y -\mu}{T}\right]^{-q/(q-1)} .
\label{tsallismu1}
\end{equation}
\indent At mid-rapidity, $y = 0$, and for zero chemical potential, as is relevant at 
the LHC, this reduces to 
\begin{equation}
\left.\frac{d^{2}N}{dp_T~dy}\right|_{y=0} = 
gV\frac{p_Tm_T}{(2\pi)^2}
\left[1+(q-1)\frac{m_T}{T}\right]^{-q/(q-1)}.
\label{tsallisfit1}
\end{equation}
\indent In the limit where the parameter $q$ goes to 1 it is well-known that this reduces to 
the standard Boltzmann distribution:
\begin{equation}
\lim_{q\rightarrow 1}\frac{d^{2}N}{dp_T~dy} = 
gV\frac{p_Tm_T\cosh y}{(2\pi)^2}
\exp\left(-\frac{m_T\cosh y -\mu}{T}\right).
\label{boltzmann}
\end{equation}
\indent The parameterization given in Eq.~\eqref{tsallismu1} is close to
the one used by various collaborations~\cite{STAR,PHENIX,ALICE,ATLAS,CMS}: 
\begin{equation}
  \frac{d^2N}{dp_T\,dy} = p_T \frac{dN}
  {dy} \frac{(n-1)(n-2)}{nC(nC + m_{0} (n-2))}
 \left[ 1 + \frac{m_T - m_{0}}{nC} \right]^{-n}  ,
\label{alice}
\end{equation}
where $n$ and $C$ are fit parameters. This corresponds to substituting~\cite{parvan} 

\begin{equation}
n\rightarrow \frac{q}{q-1}   ,
\label{n}
\end{equation}
and 
\begin{equation}
nC  \rightarrow \frac{T+m_0(q-1)}{q-1}  .
\label{nC}
\end{equation}
\indent After  this substitution Eq.~\eqref{alice} becomes
\begin{eqnarray}
  \frac{d^2N}{dp_T\,dy} =&& p_{T} \frac{{\rm d}N}
  {{\rm d}y} \frac{(n-1)(n-2)}{nC(nC + m_{0} (n-2))}\nonumber\\ 
&&\left[\frac{T}{T+m_0(q-1)}\right]^{-q/(q-1)}\nonumber\\
&&\left[ 1 + (q-1)\frac{m_T}{T} \right]^{-q/(q-1)}  .
\label{alice2}
\end{eqnarray}
\indent At mid-rapidity $y=0$ and zero chemical potential,
this has the same dependence on the 
transverse momentum as Eq.~\eqref{tsallisfit1} 
apart from an additional  factor $m_T$ on the right-hand side of Eq.~\eqref{tsallisfit1}.
However, the inclusion of the rest mass in the substitution Eq.~\eqref{nC}
is not in agreement with the Tsallis distribution as it breaks 
$m_T$ scaling which is present in Eq.~\eqref{tsallisfit1}
 but not in Eq.~\eqref{alice}. 
The inclusion of the factor $m_T$ 
leads to a more consistent interpretation of the variables $q$ and $T$.\\
\indent A very good description of transverse momenta distributions at RHIC has been
obtained in Refs~\cite{coalesce1,coalesce2} on the basis of a coalescence model 
where the Tsallis distribution is used for quarks. 
Tsallis fits have also been considered in Ref.~\cite{wong,wibig1,wibig2} but 
with a different power law leading to smaller values of the Tsallis parameter  $q$.\\
\indent Interesting results were obtained in 
Refs.~\cite{deppman,deppman3}
where spectra for identified particles were analyzed and the resulting
values for the parameters $q$ and $T$ were considered.\\

\section{Details of Transverse Momentum Distributions} 
\begin{figure}[ht]
\begin{center}
\includegraphics[width=1.1\linewidth,height=10.0cm]{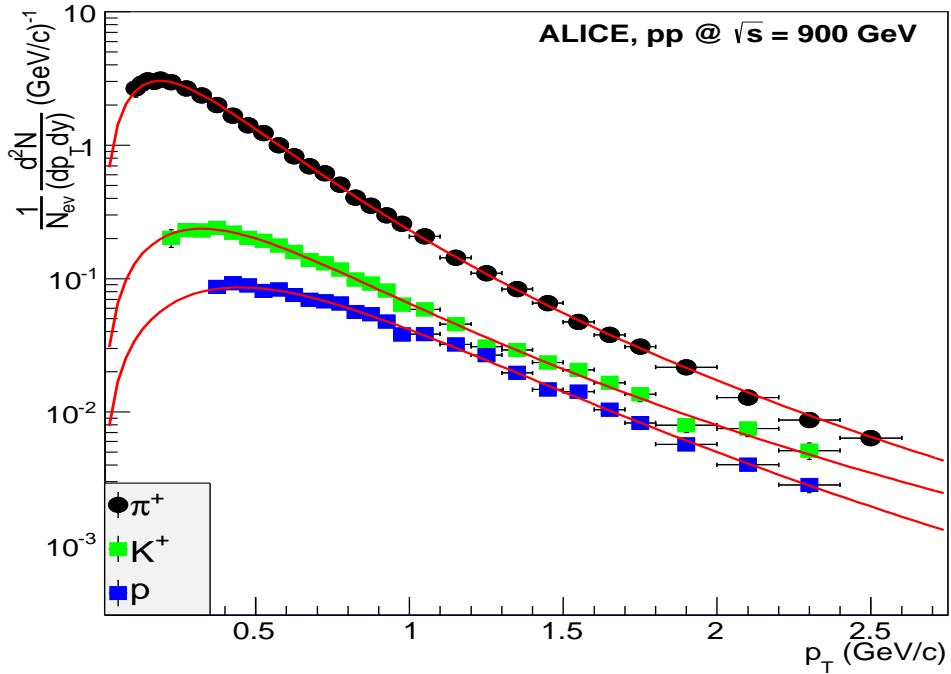}
\caption{Fits to transverse momentum distributions of positive particles~\cite{ALICE} using the Tsallis distribution.}
\label{fig:positive}
\end{center}
\end{figure}

%
\indent The transverse momentum distributions of identified particles, as obtained by the ALICE collaboration at 
900 GeV in $p-p$ collisions, are shown in \figref{fig:positive}. The fit for positive pions was made using
\begin{equation}
\left.\frac{d^{2}N}{dp_T~dy}\right|_{y=0} = 
V\frac{p_T\sqrt{p_T^2+m_\pi^2}}{(2\pi)^2}
\left[1+(q-1)\frac{\sqrt{p_T^2+m_\pi^2}}{T}\right]^{-q/(q-1)}.
\label{tsallisfitpi}
\end{equation}
with $q$, $T$ and $V$ as free parameters.\\
\\

\begin{figure}[ht]
\begin{center}
\includegraphics[width=1.1\linewidth,height=10.0cm]{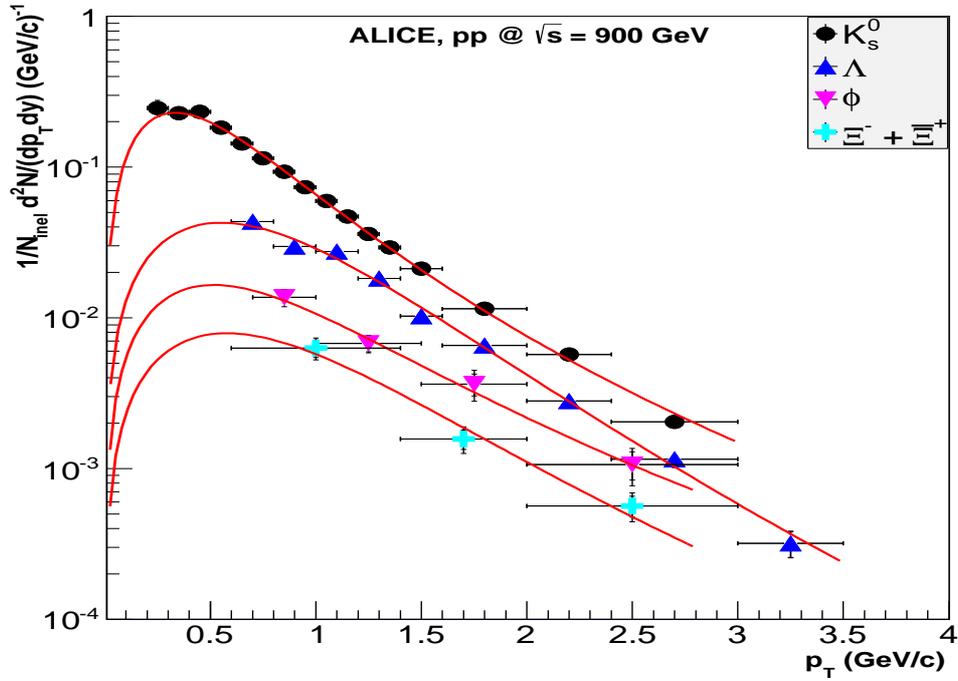}
\caption{Fits to transverse momentum distributions of strange particles~\cite{ALICE} using the Tsallis distribution.}
\label{strange}
\end{center}
\end{figure}

\begin{figure}
\centering
\includegraphics[width=1.1\linewidth,height=10.0cm]{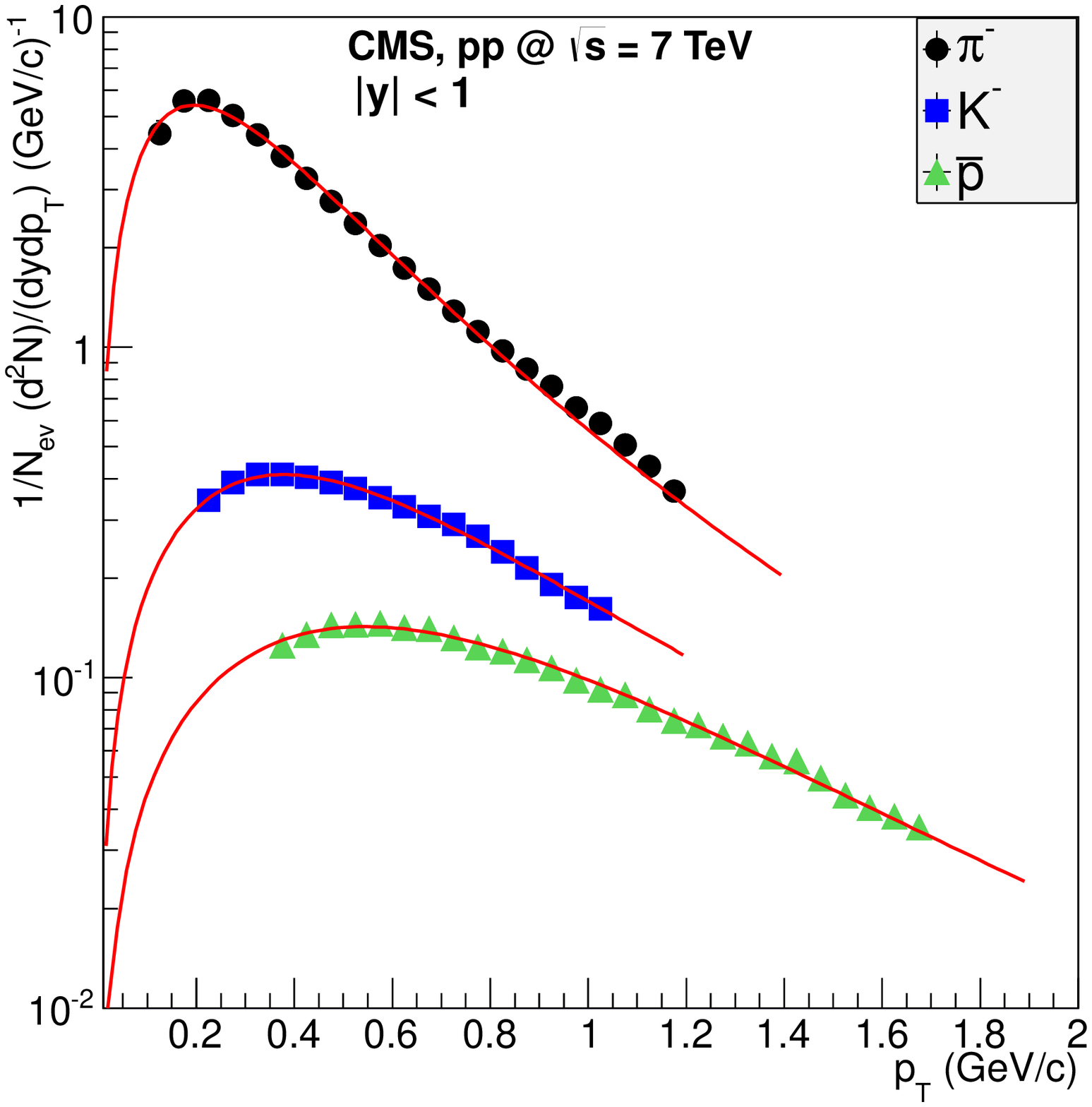}
\caption{Fits to transverse momentum distributions of negatively charged particles~\cite{CMS} using the Tsallis distribution.}
\label{cms}
\includegraphics[width=1.1\linewidth,height=10.0cm]{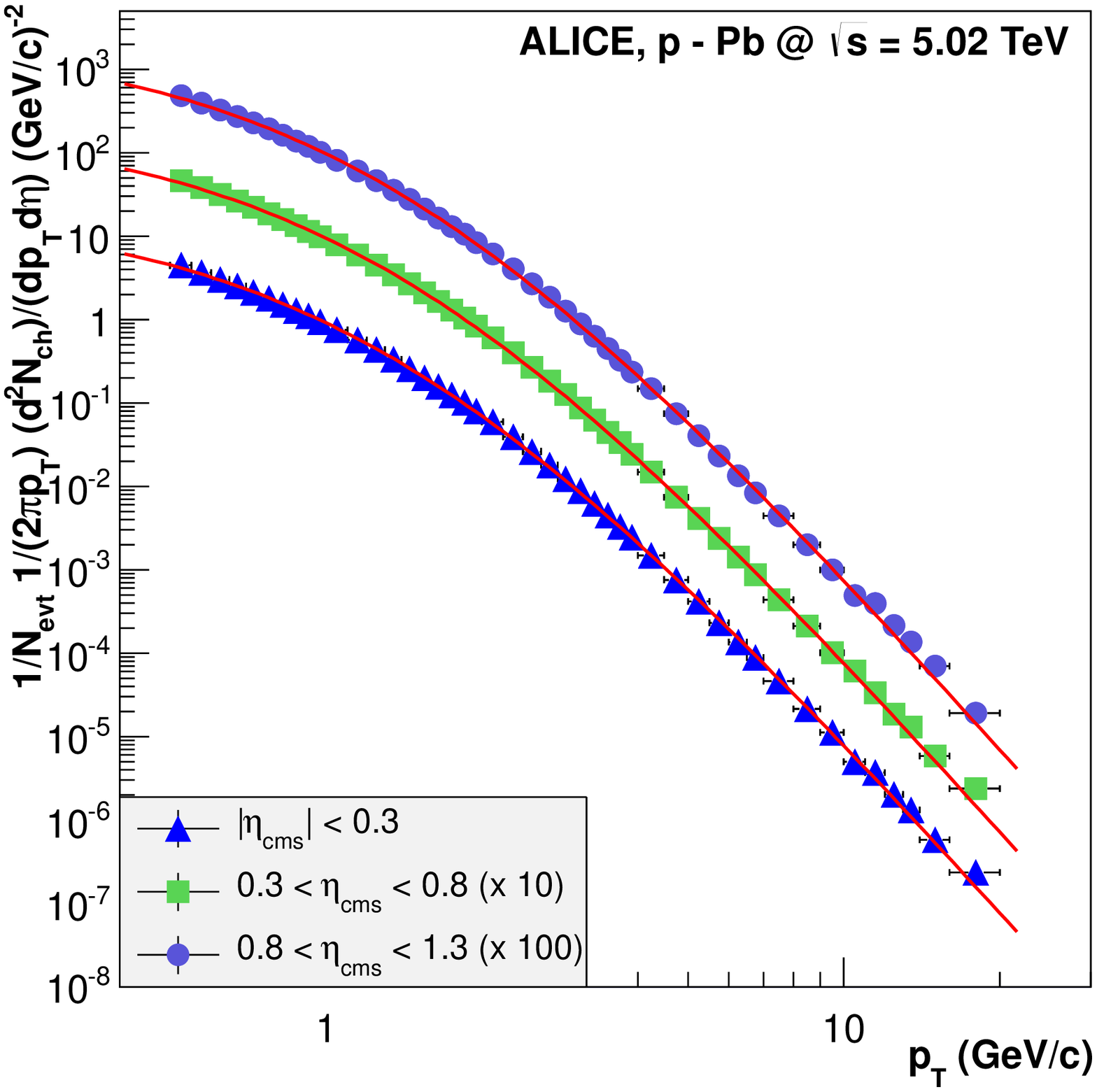}
\caption{Fits to transverse momentum distributions  in $p-Pb$ collisions obtained by the ALICE collaboration ~\cite{ALICE}
using the Tsallis distribution.}
\label{pPb}
\end{figure}

\begin{figure}[t]
\centering
\includegraphics[width=1.1\linewidth,height=10.0cm]{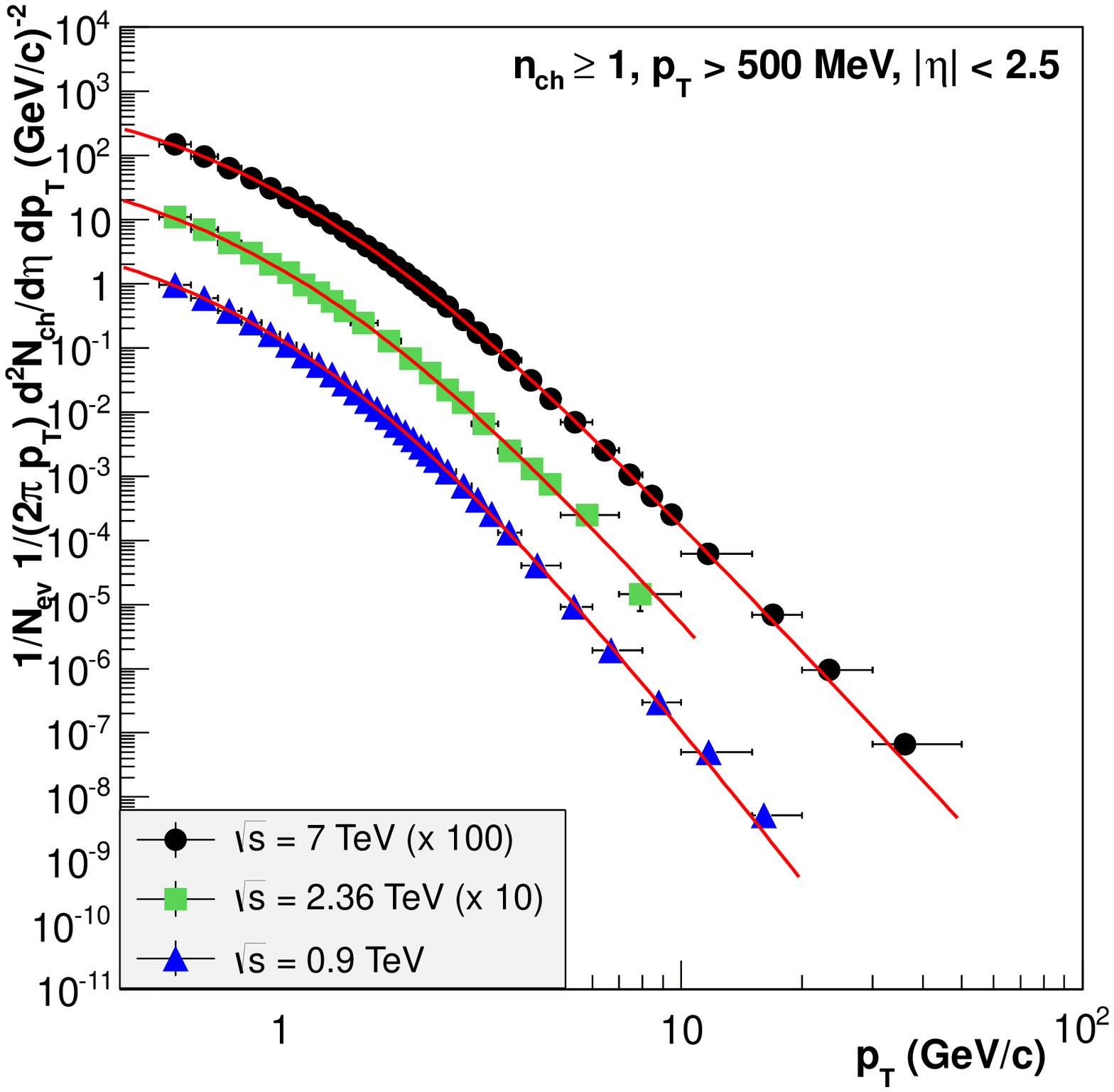}
\caption{Fits using the Tsallis distribution to transverse momentum distributions of charged 
particles measured by the ATLAS collaboration~\cite{ATLAS} in $p-p$ collisions for three  different beam energies.}
\label{chargedATLAS}
\includegraphics[width=1.1\linewidth,height=10.0cm]{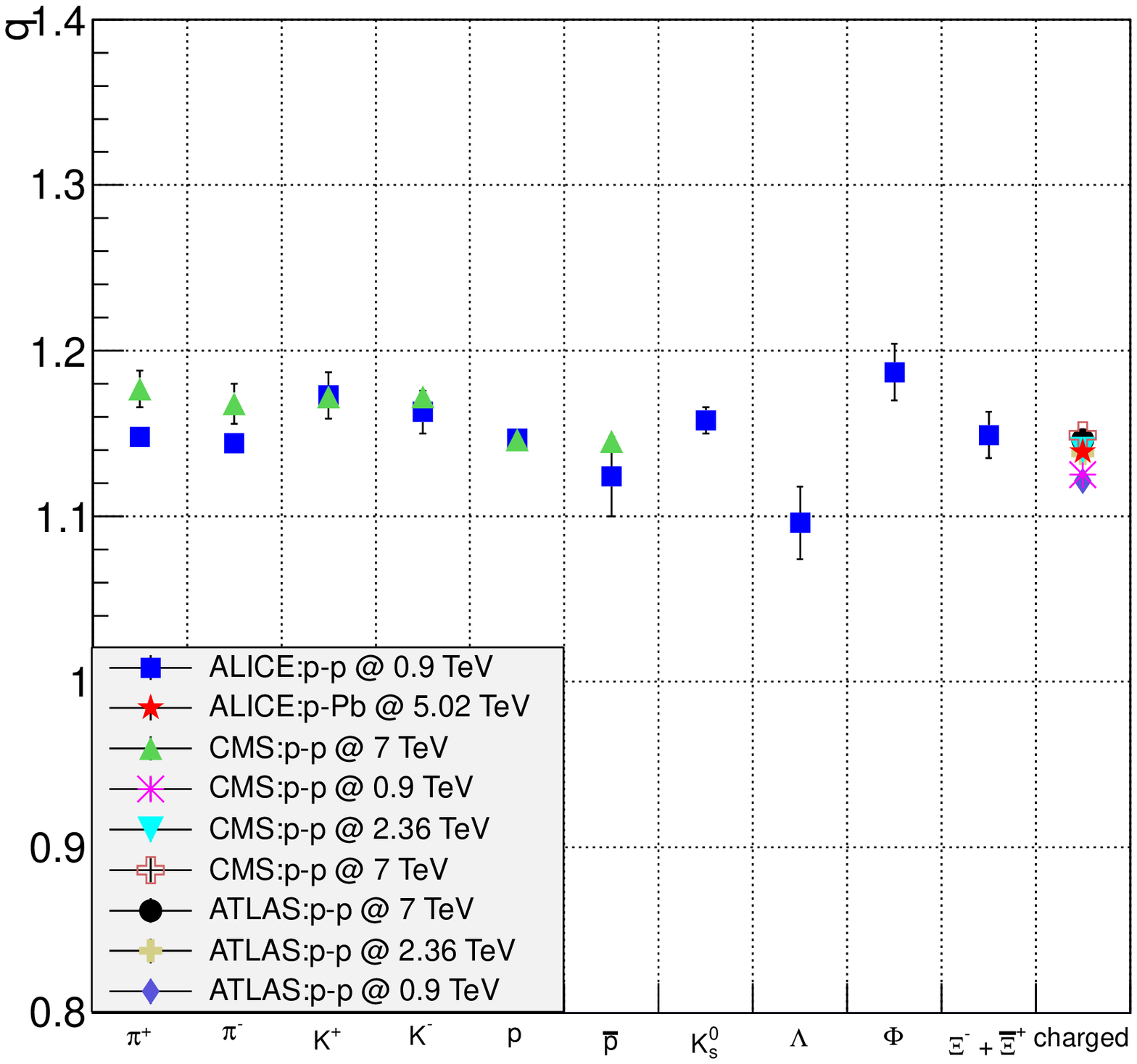}
\caption{Values of the Tsallis parameter $q$
obtained from fits to transverse momentum spectra described in the text.}
\label{azmi_q}
\end{figure}

\indent In \figref{strange} we show fits to the transverse momentum  distributions 
of strange particles obtained by the ALICE collaboration~\cite{ALICE} in $p-p$ collisions at 900 GeV.\\
\indent Similarly we show fits to the transverse momentum  distributions obtained by the CMS collaboration~\cite{CMS} in \figref{cms} and by the ATLAS collaboration in \figref{chargedATLAS}.\\
%
\indent The transverse momentum distributions of charged particles were 
fitted using a sum of three Tsallis distributions, the 
first one for $\pi^+$, the
second one for $K^+$ and the third one for protons $p$. The relative 
weights between these were 
 determined by the corresponding degeneracy factors, i.e. 1 for for $\pi^+$ and $K^+$
and 2 for  protons. 
The fit was taken at mid-rapidity and for $\mu = 0$ using the following expression was used
\begin{equation}\label{tsallisfit}
   \left.  \frac{1}{2\pi p_{T}} \frac{d^{2}N(\mathrm{charged \
particles})}{dp_{T}dy}\right|_{y=0} = \frac{2V}{(2\pi)^{3}}
\sum\limits_{i=1}^{3} g_{i} m_{T,i}
\left[1+(q-1)\frac{m_{T,i}}{T}\right]^{-\frac{q}{q-1}},
\end{equation}
where $i=(\pi^{+},K^{+},p)$  and
$g_{\pi^{+}}=1$, $g_{K^{+}}=1$ and $g_{p}=2$. The factor $2$ in front
of the right hand side of this equation takes into account the
contributions of the antiparticles $(\pi^{-},K^{-},\bar{p})$. 

%
%

%
%
%
%
The Tsallis distribution also describes the transverse momentum distributions of charged particles 
in $p-Pb$ collisions in all pseudorapidity intervals as shown in \figref{pPb}.\\

\section{Summary of Results}
The Tsallis distribution described here in Eq.~\eqref{tsallisfit1}  
leads to excellent fits to the transverse momentum distributions
in high energy $p-p$ and $p-Pb$ collisions. 
The values obtained for the Tsallis parameter $q$ are truly remarkably consistent, a feature which does not become apparent when using 
the parametrization of Eq.~\eqref{alice}.

%

%


\begin{thebibliography}{90}
\bibitem{STAR} B. I. Abelev et al. (STAR  Collaboration), Phys. Rev. C {\textbf{75}}, 064901 (2007).
\bibitem{PHENIX} A. Adare et al. (PHENIX Collaboration), Phys. Rev. C {\textbf{83}}, 052004, (2010); Phys. Rev. C {\textbf{83}}, 064903 (2011).
\bibitem{ALICE} ALICE Collaboration,  Eur. Phys. J. C {\textbf{71}} 1594 (2011); Eur. Phys. J. C {\textbf{71}} 1655 (2011); Phys. Lett. B {\textbf{693}} (2010) 53; Phys. Rev. Lett. 110 (2013) 082302.
\bibitem{ATLAS} ATLAS Collaboration, New J. Phys. {\textbf{13}} (2011) 053033.
\bibitem{CMS} CMS Collaboration, Phys. Rev. Lett. {\textbf{105}} (2010) 022002; Eur. Phys. J. C {\textbf{72}} (2012) 2164.
%
%
\bibitem{tsallis} C. Tsallis, J. Statist. Phys. {\bf 52}, 479 (1988).
\bibitem{biro} T.Bir\'o, G. Purcsel, K. \"Urm\"ossy, Eur. Phys. J. A {\textbf{40}} (2009) 325.
\bibitem{miller} J. M. Conroy, H. G. Miller, A. R. Plastino, 
Phys. Lett. A {\bf 374}, 4581 (2010).
\bibitem{worku1} J. Cleymans and D. Worku, J. Phys. G {\bf 39} (2012) 025006.
\bibitem{worku2} J. Cleymans and D. Worku, 
Eur. Phys. J. A {\bf 48} (2012) 160. 
\bibitem{coalesce1} K. \"Urm\"ossy, T.S. Bir\'{o}, Phys. Lett. B {\bf 689} (2010)  14.
\bibitem{coalesce2}  K. \"Urm\"ossy, T.S. Bir\'{o}, J. Phys. G {\bf 36} (2009)  064044.
%
%
\bibitem{wong} Cheuk-Yin Wong, G. Wilk, Acta Physica Polonica, 43 (2012) 2047.
\bibitem{wibig1} T. Wibig,  J. Phys. G: Nucl. Part. Phys. {\textbf{37}} 115009 (2010).
\bibitem{wibig2} T. Wibig, I. Kurp, JHEP {\textbf{0312}} 039 (2003).
%
\bibitem{deppman} L. Marques, E.Andrade-II, A. Deppman, arXiv:1210.1725[hep-ph]
%
%
\bibitem{deppman3} I. Sena, A. Deppman, Eur. Phys. J. A 49 (2013) 17; arXiv:1209.2367[hep-ph]
%
\bibitem{urmossy} K. \"Urm\"ossy, arXiv:1212.0260[hep-ph].
%
\bibitem{parvan}
  J.~Cleymans, G.I.~Lykasov, A.S.~Parvan, A.S.~Sorin, O.V.~Teryaev, D. Worku
  Phys.\ Lett. B {\bf 723} (2013) 351.
[arXiv:1104.0620 [hep-ph]].  
\end{thebibliography}
\end{document}